\documentclass[twocolumn,showpacs,preprintnumbers,amsmath,amssymb]{revtex4}
\usepackage{graphicx}
\usepackage{dcolumn}
\usepackage{bm}

\begin{document}
\title{
Temperature Dependence of the Chemical Potential in 
Na$_x$CoO$_2$: Implications for the Large Thermoelectric Power}

\author{Yukiaki Ishida$^1$, 
Hiromichi Ohta$^{2,3}$, Atsushi Fujimori$^1$, Hideo Hosono$^{3,4}$}

\affiliation{$^1$Department of Physics and Department of Complexity Science 
and Engineering, 
University of Tokyo,
Kashiwa, Chiba 277-8561, Japan,} 

\affiliation{$^2$Graduate School of Engineering, Nagoya University, 
Furo-cho, Chikusa, Nagoya 464-8603, Japan,} 

\affiliation{$^3$ERATO-SORST, JST, in Frontier Collaborative Research Center, 
S2-6F East, Mail-box S2-13, Tokyo Institute of Technology, 
4259 Nagatsuta-cho, Midori-ku, Yokohama 226-8503, Japan}

\affiliation{$^4$Frontier Collaborative Research Center, S2-6F East, 
Mail-box S2-13, Tokyo Institute of Technology, 4259 Nagatsuta-cho, Midori-ku, 
Yokohama 226-8503, Japan}

\date{\today}

\begin{abstract}
We have performed a temperature-dependent photoemission study of a 
Na$_x$CoO$_2$ ($x$$\sim$0.8) epitaxial thin film prepared by the 
reactive solid-phase epitaxy method. 
The chemical potential shift as a function of temperature was derived 
from the Co 3$d$ peak shift, and revealed a crossover 
from the degenerate Fermion state 
at low temperatures to the correlated hopping state 
of Co$^{3+}$/Co$^{4+}$ 
mixed-valence at high temperatures. 
This suggests that the large thermoelectric power 
at high temperatures should be considered in the correlated 
hopping picture. 
\end{abstract}



\maketitle
Thermoelectric (TE) materials directly convert heat into 
electricity. Extensive effort has been made on search for 
efficient TE materials 
for practical applications such as Peltier refrigerators 
and TE batteries \cite{Mahan_Today, DiSalvo, Sales}. 
While heavily doped semiconductors have been considered as promising 
TE materials, the recent report on the large TE power in Na$_x$CoO$_2$ 
($x$$>$0.5, $\sim$100 $\mu$V/K at room temperature) attracted much interest 
because it also showed metallic and hence high conductivity \cite{Terasaki}. 
In conventional metals, 
one expects small TE power caused by metallic diffusion 
typically of several $\mu$V/K arising from the energy dependence of the 
conductivity around the electron chemical potential $\mu$ 
\cite{Mahan_Today, Barnard}. 
Thermodynamic properties of metallic Na$_x$CoO$_2$ ($x$$>$0.5) 
are also anomalous: it exhibits a large electronic 
specific-heat coefficient \cite{Ando} and a Curie-Weiss 
magnetic susceptibility 
\cite{Ray}, 
which can be attributed to strong electron correlation \cite{Fujimori}. 
Therefore, it should be clarified whether the origin of 
the large TE power involves strong electron correlation effects or not. 
Based on the strong correlation limit, Koshibae, Tsutsui and Maekawa 
\cite{Koshibae} explained 
the large TE using a Heikes formula to include the 
spin-orbital degeneracy of the Co$^{3+}$ and Co$^{4+}$ ions. 
On the other hand, the largeness of the TE was discussed within the 
Boltzmann transport theory \cite{Singh, Takeuchi}. 

By photoemission spectroscopy (PES), one measures 
the single-particle spectral function whose energy abcissa is 
referenced to $\mu$. 
Therefore, PES is a powerful tool not only to investigate the 
quasi-particle (QP) dynamics of materials but also the 
thermodynamic quantities like $\mu$ 
\cite{Ino, ChemPot_Fuj, Keyl1}. 
Previous PES studies on bulk Na$_x$CoO$_2$ ($x$$>$0.5)
have revealed a QP peak developing in 
the narrow energy region within $\sim$100 meV from $\mu$ 
\cite{Valla, Yang, Hasan}. 
The derived hopping amplitude $t$ of the Co 3$d$ electrons was extremely small 
($|t|$$\sim$10 meV \cite{Hasan, Yang}) compared to band-structure calculation 
($|t|$$\sim$130 meV \cite{Singh}), indicating strong correlation effect. 
Interestingly, the QP peak disappeared above $\sim$200 K 
\cite{Valla, Hasan}, where the TE becomes large. 
Beyond the $\sim$100 meV region, the spectrum becomes broadened and 
dispersionless, and a Gaussian-like peak typical of low-spin $t_{2g}$ 
electron systems is observed \cite{Yang, Mizokawa}. This is in strong 
contrast to the band-structure calculations \cite{Singh}, where 
dispersions with typical bandwidths of $\sim$1 eV extend over 
the entire valence-band region. This discrepancy 
between band theory and PES spectra is reminiscent 
of the underdoped cuprates \cite{ZXShen} and 
the colossal magnetoresistive manganites \cite{Dessau}, 
where spectral weight near $\mu$ is vanishingly 
small compared to strong incoherent structures residing at higher 
binding energies.

In this Letter, we report on the temperature dependence of the 
Co 3$d$-derived single-particle spectral function including the 
incoherent part 
to obtain the information about 
the position and temperature dependence of $\mu$, 
which would be non-trivial particularly above $T$$\sim$200\,K 
where the QP disappears \cite{Valla, Hasan}. 
Here, we have made temperature-dependent PES measurements 
on a single crystalline epitaxial thin film of 
Na$_x$CoO$_2$ ($x$$=$0.83) \cite{Ohta1}. 
Epitaxial thin film surfaces are, in general, stable compared to the 
fractured surfaces of bulk single crystals, and thus advantageous for 
surface sensitive PES measurements particularly for 
temperature cycling \cite{Okazaki}.

A Na$_{0.83}$CoO$_2$ epitaxial thin film of the size 
$\sim$5\, mm\,$\times$\,10\,mm with film thickness $\sim$150\,nm 
was prepared by the 
reactive solid-phase epitaxy (R-SPE) method. 
Details of the R-SPE method are described elsewhere \cite{Ohta1}. 
After the final step of the R-SPE method, 
namely, after having annealed the sample with a 
yttria-stabilized-zirconia (YSZ) 
coverplate with NaHCO$_3$ powder in an electric furnace, the sample was 
immediately transferred into the preparation chamber of the 
spectrometer. 
In order to recover the clean surface, 
the sample was again covered with the YSZ plate and 
annealed under the 1 atmO$_2$ 
(99.9999 \%) atmosphere for 30 min at 550$^{\circ}$C 
using a Pt annealing system. 
After the surface treatment, the preparation 
chamber was quickly vented to $\sim$1.0$\times$10$^{-7}$ Torr, the 
YSZ plate was dropped off, 
and the sample was transferred 
without exposure to air into the main chamber (base 
pressure $\sim$$1 \times 10^{-10}$ Torr). 
The Al K$\alpha$ line 
({\it h}$\nu$=1486.6 eV) and the monochromatized He I line 
({\it h}$\nu$=21.22 eV) were used as excitation sources, 
and the photoelectrons were collected 
using a Scienta SES-100 analyzer. Typical energy 
resolution was 13 meV for He I and 0.8 eV for Al K$\alpha$. 

\begin{figure}[htb]
\begin{center}
\includegraphics[width=8.0cm]{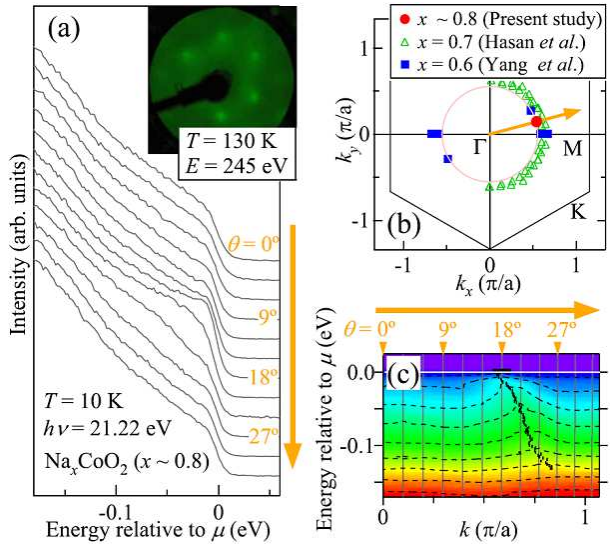}
\caption{\label{fig1} (Color) ARPES spectra of 
Na$_x$CoO$_2$ thin film prepared using the R-SPE method. 
(a) Energy distribution curves. Inset: LEED pattern for electron 
beam energy of 245 eV. 
(b) Brillouin zone of the triangular lattice. Orange arrow indicates the 
direction of the recorded dispersion. $k_F$'s determined from the present 
study and those reported by Yang {\it et al.} \cite{Yang} and 
Hasan {\it et al.} \cite{Hasan} are indicated. 
(c) Intensity plot in the $E$-$k$ plane. Red/blue 
corresponds to high/low intensity. QP dispersion is traced by 
the black dots. }
\end{center}
\end{figure}
Figure \ref{fig1}(a) shows the angle-resolved PES (ARPES) spectra of 
the Na$_x$CoO$_2$ 
thin film. 
We observed a clear dispersion from the 
$\Gamma$ point toward the Brillouin zone boundary along a cut indicated by 
an orange arrow in 
Fig.\ \ref{fig1}(b), as well as the six-fold LEED pattern shown in 
the inset of Fig.\ \ref{fig1}(a) reflecting the triangular lattice. 
A virtually carbon-free surface was confirmed using core-level PES, 
while the O 1$s$ spectra revealed a small contamination peak (5\,\% 
intensity of the O 1$s$ main peak), which was presumably due to 
adsorbed oxygen after the {\it in situ} oxygen annealing. 
These results indicate that a clean single 
crystalline surface was obtained after the R-SPE treatment. 
The QP dispersion crosses $\mu$ at $k_F$$\sim$$0.56\frac{\pi}{a}$ 
($a$: Co-Co distance) in the $E$-$k$ plane [Fig.\ \ref{fig1}(b)]. 
From comparison with the previous studies \cite{Yang, Hasan}, 
the present Fermi-surface volume is in the range of 
$\sim$0.6\,$<$\,$x$\,$<$\,$\sim$0.8 [Fig.\ \ref{fig1}(b)], 
in agreement with the value $x$$=$0.83 determined from the 
lattice constant \cite{Ohta1}. 
This indicates that the Na evaporation during the oxygen annealing was 
successfully minimized even at the outermost layer of the 
thin film by using the YSZ coverplate, and we were probing the 
electronic structure in the Curie-Weiss metal phase ($x$$>$0.5) 
showing the large TE power \cite{MLFoo}. Hereafter, we consider 
the Na concentration of the measured 
Na$_x$CoO$_2$ surface to be $x$$\sim$0.8. 

\begin{figure}[htb]
\begin{center}
\includegraphics[width=8.5cm]{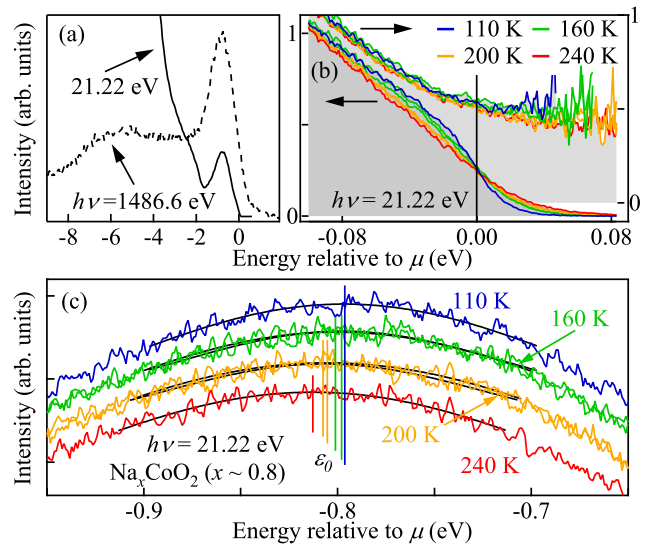}
\caption{\label{fig2} (Color) Temperature dependence of the angle-integrated 
PES spectra of Na$_{\sim}$$_{0.8}$CoO$_2$. (a) Valence-band spectra 
using photon energies of 21.22 eV and 1486.6 eV. 
(b) Temperature dependence of PES spectra in the region 
near $\mu$ (left axis). 
The temperature was cycled as 160$\to$110$\to$160$\to$200$\to$240$\to$200 
(in units of K). The spectra divided by the Fermi-Dirac function are 
also plotted (right axis). 
(c) Temperature dependence in the Co 3$d$ $t_{2g}$ peak region. 
Spectra have been offset according to the measuring temperature. 
The fitted Gaussian curves and the vertical bars indicating the peak positions 
are overlaid. }
\end{center}
\end{figure}

Next, 
we studied temperature dependence of the Co 3$d$-derived peak 
which is centered at $\sim$0.8 eV below $\mu$ as shown in Fig.\ \ref{fig2}. 
Here we note that the $\sim$0.8 eV peak was observed even when we 
increased the bulk sensitivity of 
PES using the higher excitation energy of 1486.6 eV, as shown in 
Fig.\ \ref{fig2}(a). This suggests that the broad $\sim$0.8 eV peak and 
the smallness of the QP spectral weight near $\mu$ are intrinsic 
bulk electronic properties. 
The spectra at various temperatures near $\mu$ 
are plotted in Fig.\ \ref{fig2}(b) (left axis). We have also plotted the 
spectra divided by the Fermi-Dirac function 
so as to obtain the information about the DOS 
around $\mu$. At low temperatures, 
one can see a clear Fermi cut-off, 
reflecting the metallic 
character of Na$_x$CoO$_2$ \cite{Terasaki}, and a 
negative slope in the DOS around $\mu$, which, within the 
degenerate-Fermion model, can cause an upward shift of 
$\mu$ with temperature (described later). 
The spectra show that the temperature 
dependence near $\mu$ is largely due to that 
of the Fermi-Dirac function. 
On the other hand, as shown in Fig.\ \ref{fig2}(c), the enlarged 
plot around 0.8 eV shows that 
the Co 3$d$ $t_{2g}$ peak is shifted to higher binding energies 
with temperature. 
A similar tendency was observed in the ARPES study of 
Ref.\ \cite{Hasan}. Although not so accurate, the 
temperature dependent shifts in core-level spectra (not shown) 
were also consistent with that of the Co 3$d$ peak. 
Since the temperature dependent shift of Co 3$d$ peak 
occurred reproducibly during the temperature cycling, one 
can attribute it to intrinsic changes of the electronic 
structure and not to the sample surface degradation 
during the temperature cycling. 
Here, the Co 3$d$ $t_{2g}$ 
peak positions ($\varepsilon_0$) have been determined 
by fitting the region $\sim$$-$0.8$\pm$0.1 eV to Gaussians, and are 
indicated by vertical bars. 
The deduced $T$-dependent peak shift was independent of the 
different fitted energy range. 

\begin{figure}[htb]
\begin{center}
\includegraphics[width=8.7cm]{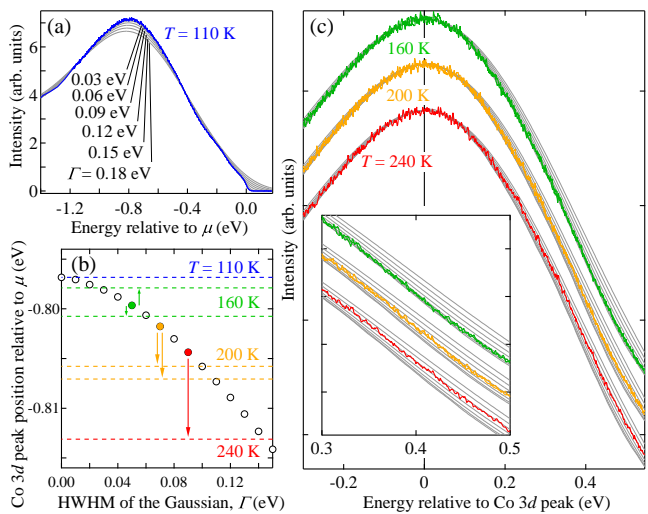}
\caption{\label{fig3} (Color) Temperature broadening and shift of the 
Co 3$d$ peak. (a) A library of Gaussian broadened 110\,K spectra. 
(b) Co 3$d$ peak position of the library spectra as a 
function of $\varGamma$ (open circles) and 
the Co 3$d$ peak positions ($\varepsilon_0$) (dashed lines). 
(c) Comparison of the line shapes of spectra with the library 
spectra. The library 
spectra for 0$<$$\varGamma$$<$0.15\,eV with 0.02-eV intervals are shown with 
thin solid curves. 
All the spectra have been normalized to the Co 3$d$ peak height, 
and the energies have been referenced to the Co 3$d$ peak position. 
Inset shows an enlarged plot of the lower binding energy side 
of the peak. }
\end{center}
\end{figure}

\begin{figure}[htb]
\begin{center}
\includegraphics[width=8cm]{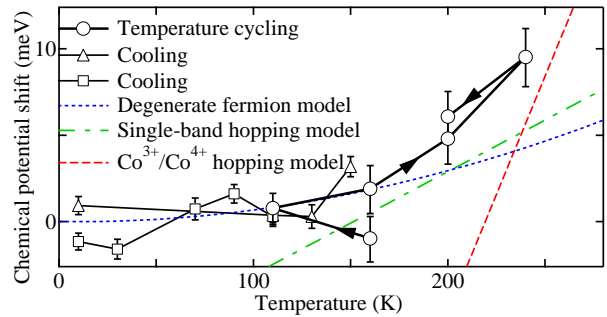}
\caption{\label{fig4} (Color) 
Chemical potential shift $\varDelta\mu$ 
of Na$_x$CoO$_2$ ($x$$\sim$0.8) derived from the PES spectra. 
$\varDelta\mu$ estimated from the degenerate 
Fermion model and the slopes of $\varDelta\mu$'s 
for the single-band hopping \cite{Chaikin} and correlated-hopping model of 
low-spin Co$^{3+}$/Co$^{4+}$ mixed-valence state \cite{Koshibae} 
are plotted for comparison (see, text). 
} 
\end{center}
\end{figure}

We shall show below that the $T$-dependent Co 3$d$ peak shift 
can be explained by a combination of a broadening-induced shift 
of the asymmetric Co 3$d$ peak and an extra $T$-dependent shift. 
We attribute the latter shift to 
the chemical potential shift, $\varDelta\mu$ 
\cite{footnote}. 
First, we show how the Co 3$d$ peak was broadened with 
temperatures. In Fig.\ \ref{fig3}(a), we have 
constructed a library of Gaussian-convoluted 110\,K spectrum. 
Since the Co 3$d$ peak was asymmetric, the peak position 
of the convoluted spectrum was shifted to lower energies 
as the half width at half the maximum ($\varGamma$) of the Gaussian 
was increased [see, Fig.\ \ref{fig3}(b)]. Then, as shown in 
Fig.\ \ref{fig3}(c), in order to reproduce the line shapes, we chose 
from the library the $\varGamma$'s for the 
$T$$=$160, 200, and 240\,K spectra as 
$\varGamma$$=$0.05, 0.07, and 0.09\,eV, respectively. 
The Co 3$d$ peak position is shifted with 
such broadening since the Co 3$d$ peak is asymmetric. 
However, this broadening-induced peak shift 
[indicated by filled circles in Fig.\ \ref{fig3}(b)] was not 
enough to explain the recorded peak shift, that is, there were 
extra $T$-dependent peak shifts as indicated by arrows 
in Fig.\ \ref{fig3}(b). 
We have interpreted this extra shift as 
$\varDelta\mu$, and have plotted it in Fig.\ \ref{fig4}. 
We have also plotted the 
relative Co 3$d$ peak shifts recorded during the cooling series 
at $T$$<$150\,K. In this temperature range, the $T$-dependent 
broadening of the Co 3$d$ peak was suppressed, 
and only the $T$-broadening of the Fermi-Dirac function 
near $\mu$ was detectable. 

In conventional metals in which charge carriers are 
degenerate Fermions, finite $\frac{\partial\mu}{\partial T}$ 
arises from the 
finite slope in the DOS around $\mu$. 
$\varDelta\mu$ is estimated within this model 
using the DOS in the inset of Fig.\ \ref{fig2}(a) 
and plotted in Fig.\ \ref{fig4} (dotted line). 
One notice that the experimentally derived $\varDelta\mu$ deviates 
from the degenerate Fermion model 
above $\sim$200\,K. This temperature coincides with the temperature above 
which the QP peak disappears in the ARPES spectra \cite{Valla, Hasan}. 
The results may be interpreted as follows: Above the 
characteristic temperature $T^*$$\sim$200 K, 
the broadening of the Fermi-Dirac function $\sim$4$k_BT$ 
exceeds the width of the narrow QP band $\sim$100 meV 
\cite{Hasan, Yang}. Then, the QP's can no more be described as degenerate 
Fermions and behave as classical particles undergoing 
thermally activated correlated hopping among the 
Co sites. 
For such carriers, 
the configuration entropy density $s$ and hence 
$\frac{\partial\mu}{\partial T}$$=$$-$$\frac{\partial s}{\partial x}$, 
where $x$ is the number of electron carriers per site, 
can be calculated following Ref.\ \cite{Chaikin}. 
In the case of Co$^{3+}$/Co$^{4+}$ mixed-valence state \cite{Koshibae}, 
	\begin{equation}
	\frac{\partial\mu}{\partial T} = 
	-k_B \ln\biggl(\frac{g_3}{g_4}\frac{1-x}{x}\biggr), 
	\label{SeebeckEq} 
	\end{equation}
where $g_3$ and $g_4$ are the spin-orbital degeneracies 
of Co$^{3+}$ and Co$^{4+}$, respectively, and $x$ is the fraction of 
Co$^{3+}$ ions. 
Here, we note that the presence of 
electron-boson coupling 
does not affect Eq.\ (\ref{SeebeckEq}). 
In Fig.\ \ref{fig4}, we have plotted 
$\varDelta\mu = -k_B\ln\bigl(\frac{g_3}
{g_4}\frac{1-x}{x}\bigr)\times T + {\rm const.}$ 
for $g_3$$=$1, 
$g_4$$=$6, and $x$$=$0.8, corresponding to the low-spin 
Co$^{3+}$/Co$^{4+}$ mixed-valence state with a dashed line \cite{Koshibae}, 
and that for $g_3$$=$1, $g_4$$=$2, 
and $x$$=$0.8 with 
a dot-dashed line, which corresponds to the single-band 
hopping model \cite{Chaikin}. 
One can recognize that the experimentally derived $\varDelta\mu$ crossovers 
at $T^*$ from the degenerate Fermion model to the 
localized-electron model 
as the temperature is increased. 

The above result for $\varDelta\mu$ 
suggests that the transport 
properties should also be treated within the correlated hopping model 
in the high temperature regime $T$$>$$T^*$ \cite{Chaikin}. 
First of all, the largeness of the TE power has indeed 
been attributed to the large 
spin-orbital entropy of low-spin Co$^{3+}$/Co$^{4+}$ mixed-valence state, 
since, at high enough temperatures, the TE power (Seebeck coefficient) 
is dominated by the entropy term 
$\frac{\mu}{T}$ compared to the energy-transport term associated with 
the kinetic energies of carriers \cite{Koshibae, KoshibaePRL, Chaikin}. 
The high temperature Hall coefficient 
also shows anomalous increase with $T$ 
\cite{MLFoo, AnomalousHall}, and its origin was discussed 
in terms of the hopping effect on the triangular lattice of 
Co$^{3+}$/Co$^{4+}$ \cite{Shastry, AnomalousHall}. 
On the other hand, the suppressed $\varDelta\mu$ at the 
low temperature regime $T$$<$$T^*$ 
indicate that the carriers are degenerate, and hence 
Boltzmann type transport for degenerate Fermion model may become valid 
\cite{Takeuchi, Singh}. 
Nevertheless, the novel 
magnetic-field suppression of the TE power is observed 
already at $T$=2.5 K 
\cite{SpinEntropy}. DC resistivity indicates anomalously 
large electron-electron scattering rate 
at $T$$<$1 K, which is also 
suppressed in a magnetic field \cite{KadowakiWoods}. 
These observations indicate that the degenerate Fermion excitations 
below $T^*$ has another low energy/temperature scale ($\ll$200 K) 
influenced by strong correlation. 

In conclusion, the experimentally derived chemical potential 
shift as a function of temperature in Na$_x$CoO$_2$ 
revealed a crossover 
from the low-temperature degenerate Fermion state to 
the high-temperature correlated hopping state involving the spin-orbital 
degeneracy of the Co$^{3+}$/Co$^{4+}$ mixed-valence at a 
characteristic temperature $T^*$$\sim$200 K. 
The anomalous transport properties at high temperatures 
such as the large TE power and the 
unsaturated Hall coefficient with temperatures 
should thus be treated within the 
correlated hopping model. 
The present work has demonstrated 
that the temperature-dependent chemical 
potential shift can be used as a measure of correlation effects in 
orbitally degenerate systems. 

We thank K.~Sugiura for collaboration, 
T.~Mizokawa, K.M.~Shen and K.~Okazaki for 
discussion, 
K.~Tanaka, H.~Yagi, H.~Wadati, M.~Hashimoto, 
M.~Takizawa and M.~Kobayashi 
for help in experiment. This work was supported 
by a Grant-in-Aid for Scientific Research in 
Priority Area (16204024) from 
the Ministry of Education, 
Culture, Sports, Science and Technology, 
Japan.


\begin{thebibliography}{99}
\bibitem{footnote}
Since the $t_{2g}$ orbital is nearly filled, the Co 3$d$ peak 
can be approximately regarded as a shallow core level 
being a good measure of $\mu$.
\end{thebibliography}
\end{document}